# Fano resonance of radiative direction in Kretschmann configuration induced by strong coupling

Gang Song,∗ Pengfei Zhang,† Gaoyan Duan, and Li Yu

School of Science, Beijing University of Posts and Telecommunications, Beijing 100876, China

Corresponding authors: songgangbupt@163.com, pf-zhang@163.com

Abstract: We investigate the strong coupling phenomenon in Kretschmann configuration involving molecular J-aggregates by using surface plasmon polaritons (SPPs) as a light source. SPPs are injected in molecular J-aggregates and propagate along the interface between molecular J-aggregates and the metal film. Both the classical method and Finite difference time domain method are used to show the radiative characteristics of light converted from SPPs in our proposed structure, with the results from the two methods being consistent with each other. Our results show that the radiative angle versus the wavelength shows a Fano line shape and the conversion efficiency of SPPs into light exhibits the strong coupling phenomenon in form of Rabi splitting. Our structure is quite sensitive to the wavelength, which has potential applications in sensing, lasing, optical switching, and nonlinear slow-light devices.

**I.INTROCUTION**

Strong interactions between cavities and quantum emitters open a new branch to manipulate light, which has many applications such as quantum networks [1], single-atom lasers [2], ultrafast single-photon switches [3] and quantum information processing [4–6]. During these decades, a series of plasmonic cavities are widely used instead of the resonant cavities to realize the strong coupling phenomenon in both theory and experiment at room temperature, which would lead the cavity and emitter mode hybridization and Rabi splitting [7], including metallic nanoparticle/metallic nanoparticle array [8–14], metallic gratings [15, 16], subwavelength holes/hole arrays [17–19], and the most popular structures of Kretschmann configurations [20–22].

Kretschmann configuration is a famous structure to generate surface plasmon polaritons (SPPs). Kinds of Kretschmann configurations are widely used in nanotweezers [23], the four-wave mixing [24], the optical bistability [25], and the strong coupling [26, 27]. All these structures need incident light injected into the structure through the prism to launch the SPPs wave. If SPPs are launched by grating or extremely focused laser and are used as a light source propagating along the interface, interesting properties in the radiative light will be expected.

In this paper, we investigate Fano resonance of the radiative direction in Krestchmann configuration involving molecular J-aggregates based on strong coupling by using classical methods. Our proposed Krestchmann configuration is made up of a silver film sandwiched between the prism and molecular J-aggregates. SPPs are injected and propagate along the interface between molecular J-aggregates and Ag film. Both the classical method and Finite Difference Time Domain (FDTD) method are used to show the radiative characteristics of the proposed structure. The parameters of molecular J-aggregates and Ag film are changed in order to show the impact on the output light, such as the radiative angle and the conversion efficiency from SPPs into light versus the wavelength. Our structure is quite sensitive to the wavelength, which has potential applications in sensors, lasing, switching, and nonlinear and slow-light devices.

## II. CALCULATION MODELS AND RESULTS

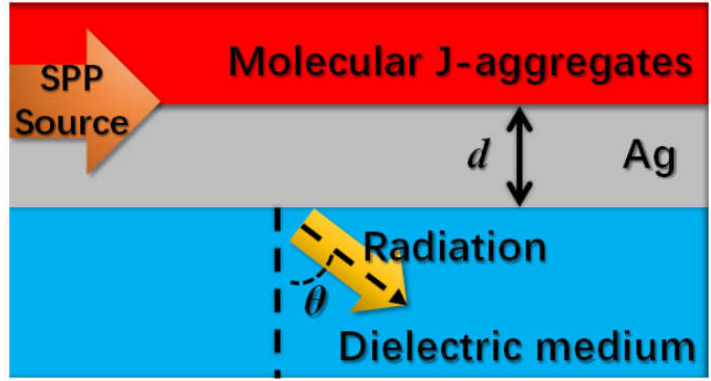

FIG. 1. (Color online). The sketch of our calculation model.

The sketch of our calculation model is shown in Fig. 1. An Ag film with the thickness of d is sandwiched between the prism and molecular J-aggregates. A SPP light source is injected and propagates along the interface between molecular J-aggregates and Ag film, which could be achieved in experiments in the way of the near field excitation or the grating configuration. The radiative angle θ in the prism can be calculated in the classical method or detected in the far field calculation by FDTD method. The dielectric constant function ($\epsilon_2$) of Ag is obtained in the Ref. [28]. The index of the prism ($\epsilon_2^{1/2}$) is chose as $3.6^{1/2}$. The molecular J-aggregates was modeled by a single Lorentzian oscillator [8, 11]:

$$\varepsilon_3 = \varepsilon_\infty + \frac{f\omega_x^2}{\omega_x^2 - \omega^2 - i\gamma_x\omega} \tag{1}$$

where, $\epsilon_\infty$ comes from the bound electron permittivity, $\omega_x$=3.22×10$^{15}$[rad/s] is the oscillator frequency, $\gamma_x$=2.45×10$^{13}$ [rad/s] is the damping constant [8]. $f$ is the oscillator strength and related with the density of molecular J-aggregates.

Usually in Kretschmann configurations, the conversion efficiency from incident light to SPPs is evaluated by the light extinction in the prism without taking account of the dissipation of SPPs propagating. Reversely, we can use the same classical method to characterize the conversion efficiency from SPPs to light. By solving the Fresnel formulas applied in the layered structure for linear media, the dimensionless intensities are obtained as [25]:

$$U_i = \frac{U_t}{4}\left|\frac{\cos k_2 d}{1+U_t}\left[(1-i\beta_{12}\tan k_2 d)+i\beta_{13}\frac{1-U_t}{1+U_t}(1-i\beta_{12}\tan k_2 d)\right]\right|^2 \quad (2a)$$

$$U_r = \frac{U_t}{4}\left|\frac{\cos k_2 d}{1+U_t}\left[(1+i\beta_{12}\tan k_2 d)+i\beta_{13}\frac{1-U_t}{1+U_t}(1+i\beta_{12}\tan k_2 d)\right]\right|^2 \quad (2b)$$

where, $U_t$ is the dimensionless intensity in molecular J-aggregates, $U_i$ and $U_r$ are the dimensionless incident and reflected intensities in the prism, $\theta_{SPPs}$ is an optimal incident angle for SPPs excitation, $k_0$ is the vacuum wave vector, $k_i^2 = k_0^2(\epsilon_i - \epsilon_1 \sin^2\theta_{SPPs})$, $i = 1, 2$, $\beta_{ij} = \epsilon_i k_j/\epsilon_j k_i$, $k_3 = k_0 g$, $i,j = 1, 2, 3$, $g^2 = \epsilon_1 \sin^2\theta_{SPPs} - \epsilon_3$, respectively.

Considering our structure, the output light comes from the decoupling from SPPs to light in the form of leaking radiation. The radiative angle θ is θSPPs, which is written as [29]:

$$\sin\theta = \sin\theta_{SPPs} = \left|\sqrt{\frac{\varepsilon_2\varepsilon_3}{\varepsilon_2+\varepsilon_3}}\varepsilon_1^{-1/2}\right| \quad (3)$$

and, the conversion efficiency from SPPs into light ($\eta$) of our structure is related with the extinction efficiency ($\eta \propto 1 - U_r/U_i$).

First, we calculate the radiative angle $\theta$. $\epsilon_\infty$ is adopted as 1.44 and $f$ is chosen as 0.005. The radiative angle is calculated by Equ.(3) and the result is shown in Fig. 2(a). The detuning energy is denoted as $\Delta = \hbar\omega - \hbar\omega_x$, where $\omega$ is the angular frequency of light, and $\hbar$ is the Planck constant. As Fig. 2(a) shown, the line shape of the curve shows the Fano shape.

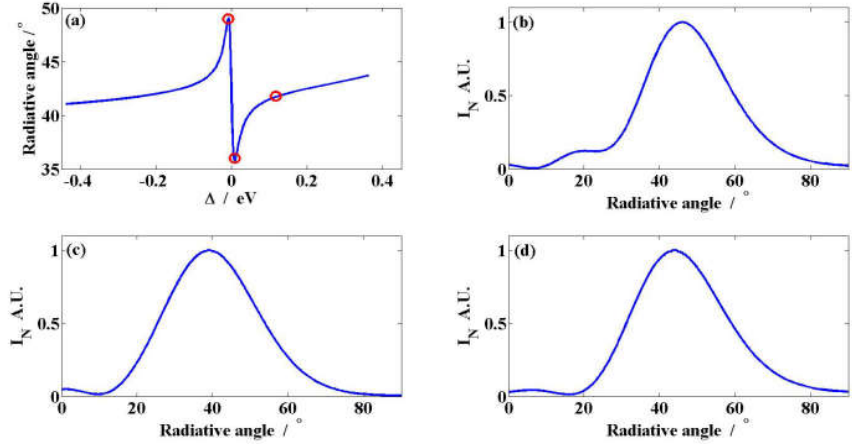

FIG. 2. (Color online). (a) The radiative angle $\theta$ versus the detuning energy $\Delta$. (b)-(d) The normalized intensity versus the radiative angle with three wavelengths: 550nm (0.1166eV), 578nm (0.007eV) and 583nm (-0.011eV), as indicated by the red circles in (a).

The mechanism behind this Fano shape is explained by the interference between the bright mode and the dark mode. SPPs propagate along the interface between molecular J-aggregate and Ag film, which shows a nonradiative mode. Molecular J-aggregates are illuminated by SPPs, which provide a bright mode. To satisfying the condition of SPPs decoupling into light, the interaction between the two modes induces a Fano line shape. Furthermore, we use Two dimension (2D) FDTD method to simulate our proposed structure, and pick up the far field information to show the radiative angle. The thickness of Ag film is adopted as 20nm ($d$=20nm), and the grid is 1nm. The length of the structure is 1000nm. The mode source is chosen as the light source, and the width of the source covers a semi-width of Ag film and the whole molecular J-aggregates. Only fundamental mode is considered. The recorder for far field information is arranged 300nm away from Ag film. We pick up three wavelengths to simulate as $\lambda$=550nm (0.1166eV), 578nm (0.007eV) and 583nm (-0.011eV), which are marked with red circles in Fig. 2(a). The curves between the radiative angle and the normalized intensity ($I_N$) are shown in the rest figures in Fig. 2, respectively.

TABLE I. Radiative angle / °

| Wavelength | Equ. (3) | FDTD |
|---|---|---|
| 550nm (0.1166eV) | 41.7° | 43,3° |
| 578nm (0.007eV) | 35.9° | 38.6° |
| 583nm (-0.011eV) | 48.5° | 45.5° |

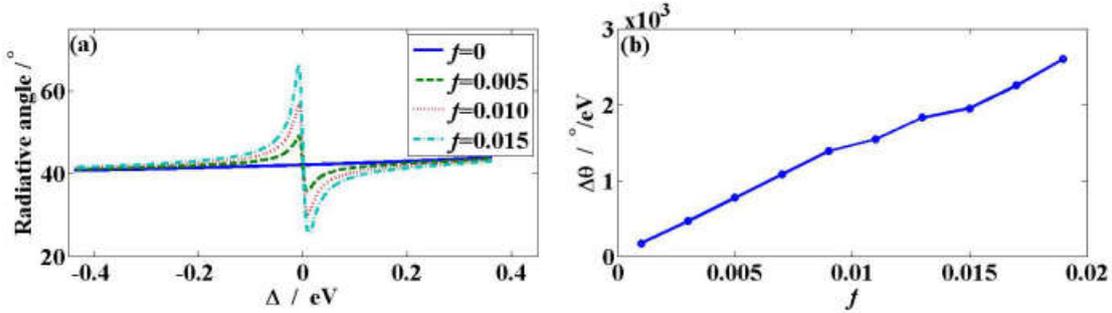

FIG. 3. (Color online). (a) Radiative angle $\theta$ versus $\Delta$ with $f$=0, 0.005, 0.01, and 0.015, respectively. (b) The on/off ratio $\Delta\theta$ versus $f$.

The radiation angles calculated from the equation and from the FDTD simulation at the three wavelengths are listed in Tab. 1. The results from these two methods match well. From Equ. (3), we may find that $\theta$ or $\theta_{SPPs}$ is not related with the thickness of Ag film. In FDTD simulations, the thickness of Ag film d is also changed and the radiative angle θ is almost unchanged. Hence, the simulation results with different d are not listed. Then, we show the impact of f on the radiative angle $\theta$. We adopt $f$=0, 0.005, 0.01, and 0.015, respectively, and use Equ. (3) to calculate θ as shown in Fig. 3(a). As shown in Fig. 3(a), there is a linear relation between the wavelength of SPPs and $\theta$ for $f$=0. The Fano line shape appears for the non-zero off. The on/off ratio ($\Delta\theta$ =($\theta_{max}$−$\theta_{min}$)/($\lambda_{SPPs;max}$-$\lambda_{SPPs;min}$) ) shows a linear relationship with $f$, where $\lambda_{SPPs}$ is the wavelength of SPPs. We further calculate $\Delta\theta$ as the function of f and plot the curve in Fig. 3(b). We find that $\Delta\theta$ increases extremely with the increase off. This is because of that the real part of the index of molecular J-aggregates changes more rapidly around $\omega_x$ with the higher $f$. If $f$ is large enough to make $\text{Re}[\epsilon_3^{1/2}]$ much larger than $\epsilon_1^{1/2}$ with $\omega$ around $\omega_x$, $\theta$ turns to be complex and $\text{Re}[\theta]$ is no less than 90°, meaning no light converted from the SPPs.

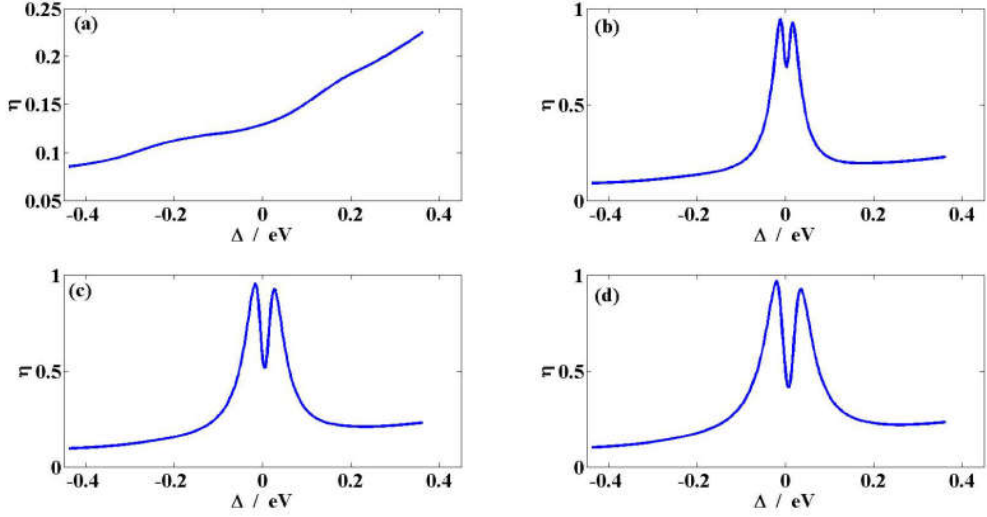

FIG. 4. (Color online). Conversion efficiency $\eta$ versus $\Delta$ with $d$=20nm and $f$=0, 0.005, 0.01, and 0.015, respectively

The influence of $f$ on the conversion efficiency $\eta$ from SPPs into light is also considered. Here, we choose the thickness of Ag film $d$ as 20nm, and the calculation results with $f$=0, 0.005, 0.01, and 0.015 are shown in Fig. 4, respectively.

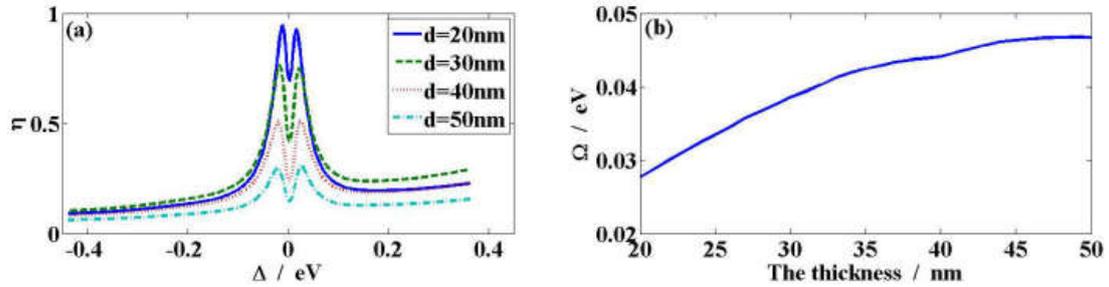

FIG. 5. (Color online).(a) Conversion efficiency $\eta$ versus $\Delta$ with $f$=0.005 $d$=20nm, 30nm, 40nm and 50nm, respectively. (b) Rabi splitting $\Omega$ versus $d$.

There are two peaks in Fig. 4(b)-4(d) with different dips, while there is no peak in Fig. 4(a). The two peaks originate from the Rabi splitting/ the strong coupling. The molecular J-aggregates don't only help to generate the strong coupling effect, but also amplify the intensity of SPPs along the interface between molecular J-aggregates and Ag film due to the image part of the dielectric constant of molecular J-aggregates. With the increase off, the dip becomes deeper and the distance between the two splitting peaks becomes larger.

For our structure, SPPs as light sources are injected into our structure. They propagate along the interface between molecular J-aggregates and Ag film and meet molecular J-aggregates. At the same time, the strong coupling between SPPs and molecular J-aggregates happens in forms of Rabi splitting. And the splitting SPPs decouple/converse into light. We obtain the dispersion relation of SPPs for multiple layer structure as following [29]:

$$e^{-2k_2 d} = \frac{k_2/\varepsilon_2 + k_1/\varepsilon_1}{k_2/\varepsilon_2 - k_1/\varepsilon_1} \frac{k_2/\varepsilon_2 + k_3/\varepsilon_3}{k_2/\varepsilon_2 - k_3/\varepsilon_3} \qquad (4)$$

$k_2$ can be simplify as:

$$k_2 = \varepsilon_2 \frac{\frac{k_1}{\varepsilon_1} + \frac{k_3}{\varepsilon_3} - \sqrt{\left(\frac{k_1}{\varepsilon_1} - \frac{k_3}{\varepsilon_3}\right)^2 - \frac{4}{\varepsilon_1 d}\left(\frac{k_1}{\varepsilon_1} + \frac{k_3}{\varepsilon_3}\right)}}{2} \qquad (5)$$

According to the analysis method in Ref.[30] and neglecting the dissipation of our structure, the two splitting frequencies are approximated as:

$$\omega_\pm = \frac{\kappa}{2} + \frac{\omega_x}{2} \pm \frac{1}{2}\sqrt{A + (k - \omega_x)^2} \qquad (6)$$

where, $A$ is related with the number density of dipole moments $(N/V)^{1/2}$ ($A \propto N/V$), and $\kappa^2 = k_2^2 n_{eff}^{-2} c^2$. Here, we predict that with d increasing, the different between the two splitting peaks $\Omega = |\omega_+ - \omega_-|$ increases based on Equ. (6). We choose four thicknesses of Ag film ($d$=20nm, 30nm, 40nm and 50nm) to calculate $\eta$, and the results are shown in Fig. 5(a). $f$ is fixed at 0.005. By using Equ. (2) and changing $d$ from 20nm to 50nm with the step of 1nm, we also obtain the Rabi splitting $\Omega$ as shown in Fig. 5(b).

As Fig. 5 shown, $\eta$ decreases with the decrease of $d$ due to the increase of the dissipation of our structure. There is a linear relation between $\Omega$ and $d$, which is predicted in Equ. (6).

**III. Summary**

In summary, we investigate the radiative angle showing a Fano line shape in Kretschmann configuration involving molecular J-aggregates, while the conversion efficiency from SPPs into light exhibits the strong coupling/Rabi splitting phenomenon. Molecular J-aggregates don't only influence on the Fano line shape, but also impact on the strong coupling effect. FDTD method is also employed to check the radiative angle, and the results of the analytical method are well matched with the one of FDTD simulations. Our structure is quite sensitive to the wavelength, which has potential applications in sensors, lasing, optical switching, and nonlinear and slow-light devices.

**Acknowledgement**

This work was supported by Ministry of Science and Technology of China (Grant No. 2016YFA0301300)